\documentclass[aps,twocolumn,prd,amsmath,floatfix,nofootinbib,preprintnumbers,superscriptaddress]{revtex4-2}
\usepackage{xcolor}
\usepackage{graphicx}
\usepackage{lineno}
\usepackage{nicefrac}
\usepackage{subfigure,slashed}
\usepackage[colorlinks = true,linkcolor = blue, urlcolor  = blue,citecolor = red,anchorcolor = blue]{hyperref}
\let\oldcite\cite
\renewcommand{\cite}[1]{\mbox{\oldcite{#1}}}

\def\traditional{{\tt traditional}}
\def\pnSRC{{\tt pnSRC}}
\def\baseSRC{{\tt baseSRC}}

\def\ncteqpp{\hbox{nCTEQ++}}

\def\chidat{\chi^2/N_{\rm data}}

\newcommand{\orcid}[1]{\,\href{https://orcid.org/#1}{\includegraphics[width=9pt]{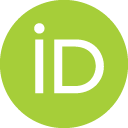}}\,}
\newcommand{\orcidTJ}{0000-0002-1334-7607} %
\newcommand{\orcidPD}{0000-0001-7960-7953} %
\newcommand{\orcidOH}{0000-0002-4890-6544} %
\newcommand{\orcidTK}{0000-0002-7516-8292} %
\newcommand{\orcidMK}{0000-0002-4665-3088} %
\newcommand{\orcidKK}{0000-0003-1412-447X} %
\newcommand{\orcidAK}{0000-0002-4090-0084} %
\newcommand{\orcidND}{0000-0003-0962-631X} %
\newcommand{\orcidJM}{0000-0001-9343-9351} %
\newcommand{\orcidFM}{0000-0002-3888-1697} %
\newcommand{\orcidFO}{0000-0001-6799-2436} %
\newcommand{\orcidIS}{0000-0003-0373-474X} %
\newcommand{\orcidJY}{0000-0001-8366-0968} %
\newcommand{\orcidPR}{0000-0002-8570-5506} %
\newcommand{\orcidRR}{0000-0002-3316-2175} %
\newcommand{\orcidEP}{0000-0001-9058-2590} %
\newcommand{\orcidAD}{0009-0009-4902-6408} %

\usepackage{ulem}

\newcommand{\sectionX}[1]{\vspace{6pt}\noindent \textbf{#1:} }
\newcommand{\subsectionX}[1]{\vspace{6pt}\noindent \textbf{#1:} }
\begin{document}

\preprint{MS-TP-22-13}
\preprint{IFJPAN-IV-2022-21}
\preprint{SMU-HEP-23-08}
\preprint{FNAL-PUB-23-146-ND}

\def\mit{\affiliation{Massachusetts Institute of Technology, Cambridge, Massachusetts 02139, USA}}
\def\smu{\affiliation{Department of Physics, Southern Methodist University,
    Dallas, TX 75275-0175, U.S.A.}}
\def\jlab{\affiliation{Jefferson Lab, Newport News, VA 23606, U.S.A.}}
\def\kit{\affiliation{Institute for Theoretical Physics, KIT,  D-76131  Karlsruhe, Germany}}
\def\krakow{\affiliation{Institute of Nuclear Physics Polish Academy of Sciences, PL-31342 Krakow, Poland}}
\def\muenster{\affiliation{Institut f{\"u}r Theoretische Physik, Universit{\"a}t M{\"u}nster,             \\Wilhelm-Klemm-Stra{\ss}e 9, D-48149 M{\"u}nster, Germany}}
\def\lpsc{\affiliation{Laboratoire de Physique Subatomique et de Cosmologie, Université Grenoble-Alpes, 
    \\CNRS/IN2P3, 53 avenue des Martyrs, 38026 Grenoble, France}}
\def\hampton{\affiliation{Hampton University, Hampton, VA 23668, USA}}
\def\fnal{\affiliation{Fermi National Accelerator Laboratory, Batavia, Illinois 60510, USA}}
\def\iit{\affiliation{Department of Physics, Illinois Institute of Technology, Chicago, Illinois 60616, USA}}
\def\fsu{\affiliation{Department of Physics, Florida State University, Tallahassee, Florida 32306-4350, USA}}
\def\wales{\affiliation{School of Physics, The University of New South Wales, Sydney NSW 2052, Australia}}
\def\jyv{\affiliation{University of Jyväskylä, Department of Physics, P.O.\ Box 35, FI-40014 University of Jyväskylä, Finland}}
\def\helsinki{\affiliation{Helsinki Institute of Physics, P.O.\ Box 64, FI-00014 University of Helsinki, Finland}}
\def\anl{\affiliation{High Energy Physics Division, Argonne National Laboratory, Argonne, Illinois 60439, USA}}
\def\fzj{\affiliation{Institut für Energie- und Klimaforschung, Forschungszentrum Jülich GmbH, 52425 Jülich, Germany}}
\def\telaviv{\affiliation{School of Physics and Astronomy, Tel Aviv University, Tel Aviv 6997845, Israel}}

\title{Modification of Quark-Gluon Distributions in Nuclei 
by Correlated Nucleon Pairs}
\author{A.W.~Denniston\orcid{\orcidAD}}
\email{awild@mit.edu}
\mit{}
\author{T.~Je\v{z}o\orcid{\orcidTJ}}
\email{tomas.jezo@uni-muenster.de}
\muenster{}
\author{A.~Kusina\orcid{\orcidAK}} 
\krakow{}
\author{N. Derakhshanian\orcid{\orcidND}} 
\krakow{}
\author{P.~Duwent\"aster\orcid{\orcidPD}}
\muenster{}
\jyv{}
\helsinki{}
\author{O.~Hen\orcid{\orcidOH}} 
\mit{}
\author{C.~Keppel\orcid{\orcidTK}} 
\jlab{}
\author{M.~Klasen\orcid{\orcidMK}}
\muenster{}
\wales{}
\author{K.~Kova\v{r}\'{i}k\orcid{\orcidKK}} 
\muenster{}
\author{J.G.~Morf\'{i}n\orcid{\orcidJM}} 
\fnal{}
\author{K.F.~Muzakka\orcid{\orcidFM}}
\muenster{}
\fzj{}
\author{F.I.~Olness\orcid{\orcidFO}} 
\smu{}
\author{E.~Piasetzky\orcid{\orcidEP}} 
\telaviv{}
\author{P.~Risse\orcid{\orcidPR}}
\muenster{}
\author{R.~Ruiz\orcid{\orcidRR}}
\krakow{}
\author{I.~Schienbein\orcid{\orcidIS}} 
\lpsc{}
\author{J.Y.~Yu.\orcid{\orcidJY}}
\lpsc{}

\begin{abstract}
    
We extend the QCD Parton Model analysis using a factorized nuclear structure
model incorporating individual nucleons and pairs of correlated nucleons. Our
analysis of high-energy data from lepton Deep-Inelastic Scattering, Drell-Yan
and W/Z production simultaneously extracts the universal effective distribution
of quarks and gluons inside correlated nucleon pairs, and their
nucleus-specific fractions.  Such successful extraction of these universal
distributions marks a significant advance in our understanding of nuclear
structure properties connecting nucleon- and parton-level quantities.

\end{abstract}

\date{\today}
\maketitle

\sectionX{Introduction}
Subatomic systems, such as nucleons and atomic nuclei, and dense
astrophysical matter, derive their properties from the many-body
interactions between their constituent quarks and gluons
which  are described by the theory
of Quantum Chromodynamics (QCD). The strongly coupled nature of QCD
links the momentum distribution of quarks and gluons in these systems
to their structure and emergent properties, such as mass and spin~\cite{AbdulKhalek:2021gbh}.
Therefore, it is crucial to understand the momentum distribution of quarks and
gluons in nucleons and nuclei.

By analyzing high-energy collisions between leptons, nucleons, and
nuclei, using a well-established QCD factorization formalism~\cite{Collins:1987pm,Collins:1998rz,Collins:2011zzd},
parton distribution
functions (PDFs) can be extracted for both nucleons and
nuclei. These PDFs describe the fraction of the total system momentum
carried by different flavored quarks 
and gluons~\cite{Bailey:2020ooq,NNPDF:2021njg,Hou:2019efy}.
The nuclear PDFs (nPDFs) are observed to differ from the simple
sum of their free-proton and free-neutron PDFs, indicating a measurable
role of nuclear dynamics that remains to be understood~\cite{Kovarik:2015cma,Eskola:2021nhw,Helenius:2021tof,AbdulKhalek:2022fyi,Segarra:2020gtj,Ruiz:2023ozv}.

Compared to the nucleon PDFs, the nuclear PDFs are impacted by various
nuclear effects due to the spatial distribution of nucleons in the
nucleus (shadowing and anti-shadowing), their motion (Fermi-motion),
and strong interactions between nucleons affecting their internal
parton structure~\cite{Alvioli:2009iw,Kopeliovich:2012kw}. 
The standard
global analyses (such as nCTEQ~\cite{Kovarik:2015cma,Kusina:2020lyz,Segarra:2020gtj,Duwentaster:2021ioo,Duwentaster:2022kpv,Muzakka:2022wey}, 
EPPS~\cite{Eskola:2016oht,Eskola:2021nhw}, nNNPDF \cite{AbdulKhalek:2019mzd,AbdulKhalek:2020yuc,AbdulKhalek:2022fyi}, 
TUJU~\cite{Walt:2019slu,Helenius:2021tof}, DSSZ~\cite{deFlorian:2011fp}, Khanpour et al.~\cite{Khanpour:2016pph,Khanpour:2020zyu}) extract parton densities
inside the {\it full nucleus} using experimental data.

Here we propose to use state-of-the-art knowledge of nuclear 
theory to guide the development of nuclear PDFs:

Nuclei are commonly described as a group of independent nucleons
moving within an effective average mean field that leads to the
population of atom-like nuclear shells~\cite{Caurier:2004gf}. 
In this picture, nuclear effects are modeled in the nPDFs 
by consistently
modifying all nucleons under the effective influence of the nuclear mean-field.
It is important to note that this does not 
allow for a meaningful interpretation in terms of modified parton
densities inside bound single nucleon states since they incorporate
effects from many nucleon states present in the data.

Nuclear structure studies show that
the formation of short-lived excitations, caused by
strongly-interacting Short Range Correlated (SRC) nucleon pairs~\cite{Subedi:2008zz,Hen:2016kwk,Atti:2015eda,CiofiDegliAtti:1989eg} 
are significant.
While the abundance of SRC pairs differs among nuclei, they are
predominantly proton-neutron pairs~\cite{Piasetzky:2006ai,Subedi:2008zz,LabHallA:2014wqo,Hen:2014nza,Duer:2018sxh,Korover:2020lqf,West:2020tyo}
and have the same behavior in all nuclei~\cite{Hen:2016kwk,Atti:2015eda,Cohen:2018gzh,CLAS:2020mom,CiofidegliAtti:2017xtx}.
Consequently, SRC pairs have universal properties across nuclei and
typical separation energies of 15\% -- 30\% of the nucleon mass, which is
significantly higher than that of mean-field states~\cite{CLAS:2020mom,Weiss:2018tbu,CiofidegliAtti:1995qe,CiofidegliAtti:2017xtx}. 
The large
energies and significant spatial overlaps of SRC pairs motivated
various studies of the relation between SRC pairs and bound-nucleon
structure~\cite{Hen:2016kwk}. 

This analysis studies nPDFs based on our understanding of
high-resolution nuclear structure with SRCs. 
It allows for the first time to split the partonic
structure inside nuclei into mean-field and SRC contributions and to
extract information on nuclear structure from an analysis of the
partonic content of nuclei.
We try to take a model-agnostic
approach by focusing on the broad-scale features common to modern
high-resolution nuclear structure models, minimizing dependence on
specific model details. 

\sectionX{Short-Range Nuclear Structure}
The fundamental quantity of nuclear structure that is relevant for our study is the nuclear spectral function $S_A(k,E)$ that
defines the probability of finding a nucleon with momentum \textit{k} and separation energy \textit{E} in a nucleus with
mass number \textit{A}. We use a normalization convention of:
$\int S_A(k,E) k^2 dk dE \equiv 1$.

Direct many-body calculations of  $S_A(k,E)$ are computationally unfeasible for $A>3$ nuclei.
Therefore, we employ an established approximation where 
the spectral function is divided into two parts~\cite{Atti:2015eda}:
\begin{eqnarray}
  S_A(k,E) &=&
  S_A^{\mathrm{MF}}(k,E)+S_A^{\mathrm{SRC}}(k,E) \quad ,
\label{eq:1specf}
\end{eqnarray}
with  $S_A^{\mathrm{MF}}(k,E)$ representing the single nucleons in a mean-field (MF),
and 
$S_A^{\mathrm{SRC}}(k,E)$ representing the spectral function of nucleons in SRC pairs.

The separation presented in Eq.\,(\ref{eq:1specf}) is rooted in the vastly different energy scales associated with the single-nucleon
mean-field potential and the interaction energy inside SRC pairs. While mean-field nucleons have momenta and energy below
nuclear Fermi momentum ($k_F \sim  250$~MeV/c) and Fermi energy $E_F \sim 35$~MeV, the strong
pair-wise interaction energy inside SRC pairs leads to relative momenta of 300 $\sim$ 800 MeV/c and separation
energies of 150 -- 400 MeV~\cite{CLAS:2020mom,Weiss:2018tbu,CiofidegliAtti:1995qe,CiofidegliAtti:2017xtx}. 

The high-energy scale associated with interactions in SRC pairs leads to a further factorization of their spectral
function into a universal (nucleus independent) pair spectral function distribution, scaled by a (nucleus dependent) pair
abundance factor~\cite{Weiss:2018tbu}:
\begin{align}
 S_A^{\mathrm{SRC}}&(k,E) \approx
 \nonumber \\
  &\frac{Z}{A} \, C^A_p \times S_p^{\mathrm{SRC}}(k,E)
  +\frac{N}{A} C^A_n \times S_n^{\mathrm{SRC}}(k,E) \ .
\label{eq:2specf}
\end{align}
In the above approximation, we do consider all possible 
$(pn),\ (pp), \text{and}\ (nn)$ nucleon-nucleon pairs 
by introducing effective coefficients, $C^A_{p(n)}$, 
that sum the number of $(pn)$, $(pp)$ and $(nn)$, $(np)$ pairs, 
respectively.

Here  $C^A_{N}$ ($N{=}p,n$)
are nucleus-dependent constants that `count' the fraction of nucleons in SRC pairs, and  $S_{N}^{\mathrm{SRC}}(k,E)$  are universal (nucleus-independent) pair distributions
that are dominated by the strong nucleon-nucleon interaction at short-distance. $Z$ and $N$ are the total
number of protons and neutrons in the nucleus ($Z{+}N{=}A$).  The universal pair spectral
functions follow normalization conventions of
${\int S_{N}^{\mathrm{SRC}} (k,E) k^2 dk dE  \equiv 1}$  ($N {=} p, n$) and therefore
${\int S_A^{\mathrm{MF}}(k,E) k^2 dk dE} = {1-(ZC^A_p+N C^A_n)/A}$.

Combining Eqs.\,(\ref{eq:1specf}) and~(\ref{eq:2specf}) we obtain:
\begin{eqnarray}
  S_A(k,E)
  \approx
  S_A^{M\/F}(k,E)
&+& \frac{Z}{A} C^A_p  \times  S_p^{\mathrm{SRC}}(k,E) 
\nonumber \\
&+& \frac{N}{A} C^A_n  \times  S_n^{\mathrm{SRC}}(k,E) \ ,
\label{eq:3specf}
\end{eqnarray}
where we emphasize that the mean-field term  $S_A^{\mathrm{MF}}(k,E)$  captures low-energy, single nucleon
dynamics and the SRC terms  $C^A_N \times S_{N}^{\mathrm{SRC}}(k,E)$  captures universal high-energy
nucleon-pair dynamics. 

We note that the approximation presented in  Eq.\,(\ref{eq:3specf}) enjoys significant support~\cite{Weiss:2018tbu,Pybus:2020itv,Weiss:2015mba,Weiss:2016obx,Cruz-Torres:2019fum,CLAS:2020mom,Korover:2020lqf} by recent analyses
of ab initio many-body nuclear structure calculations and high-energy electro-induced nucleon knockout
measurements.
Furthermore,  Eq.\,(\ref{eq:1specf}) can in principle be extended to also include three-nucleon
correlation effects which are neglected in the context of this work.

\sectionX{SRC Motivated nuclear-PDFs}
nPDFs are defined within perturbative QCD using the framework of collinear factorization~\cite{Collins:1985ue,Collins:1998rz}. This framework allows the computation of cross-sections, $d\sigma_{AB\to X}$, for scattering of particles $A$, $B$ into final state $X$ as convolutions of perturbatively calculable  parton-level short-distance cross-sections, $d\hat{\sigma}_{ij\to X}$, and non-perturbative PDFs, $f_{i(j)}$, 
where $i$ and $j$ sum over the partonic content of hadrons $A$ and $B$, respectively.

Introducing these nuclear quark and gluon distributions to the nuclear structure model of 
Eqs.~(\ref{eq:1specf}) and~(\ref{eq:2specf}) leads to an nPDF parametrization that is comprised of a linear combination of free-nucleon PDFs (representing the quasi-free-nucleons), 
and SRC PDFs which describe the universal quark and gluon distributions inside an SRC pair:
\begin{align}
 f_i^A&(x,Q) = 
\nonumber \\
  &\frac{Z}{A}
  \left[ (1-C_p^A)  \times  f_i^p(x,Q) +C_p^A  \times  f_i^{\mathrm{SRC}\, p}(x,Q) \right] 
  \nonumber \\
  +\,\,\,
  &\frac{N}{A} \Big[ (1-C_n^A)  \times  f_i^n(x,Q) + C_n^A  \times  f_i^{\mathrm{SRC}\, n}(x,Q)  \Big] .
  \nonumber \\ 
\label{eq:4fSRC}
\end{align}
Here, $f_i^A(x,Q)$ is the nPDF of parton type  $i$ (gluon or quark flavors) in a nucleus with mass number
{$A$}, carrying momentum fraction $x$ at energy scale $Q$.  $f_i^N(x,Q)$ and 
$f_i^{\mathrm{SRC}\, N}(x,Q)$ are the PDFs of the free-nucleon and of the modified nucleon in an SRC pair, respectively. 
Here we implicitly assume that $f_i^{\mathrm{SRC}\, N}(x,Q)$ can be defined via a collinear factorization framework, 
and that the proton and neutron distributions are related by isospin.
Therefore, we apply the tools from perturbative QCD used for free-nucleon PDFs to arrive at the physical predictions ({\it e.g.}, DGLAP evolution).

A key feature of this framework is that the nuclear structure dependence is fully encapsulated in the fraction of nucleons in SRC pairs $C^A_N$.

As we do not separate the individual effects of proton-proton, neutron-neutron
and proton-neutron SRCs,  Eq.\,(\ref{eq:4fSRC}) relates  the modified structure of an average nucleon in an SRC pair, independent of
its pair configuration.

\begin{figure}[tb]
\centering
\includegraphics[width=0.45\textwidth]{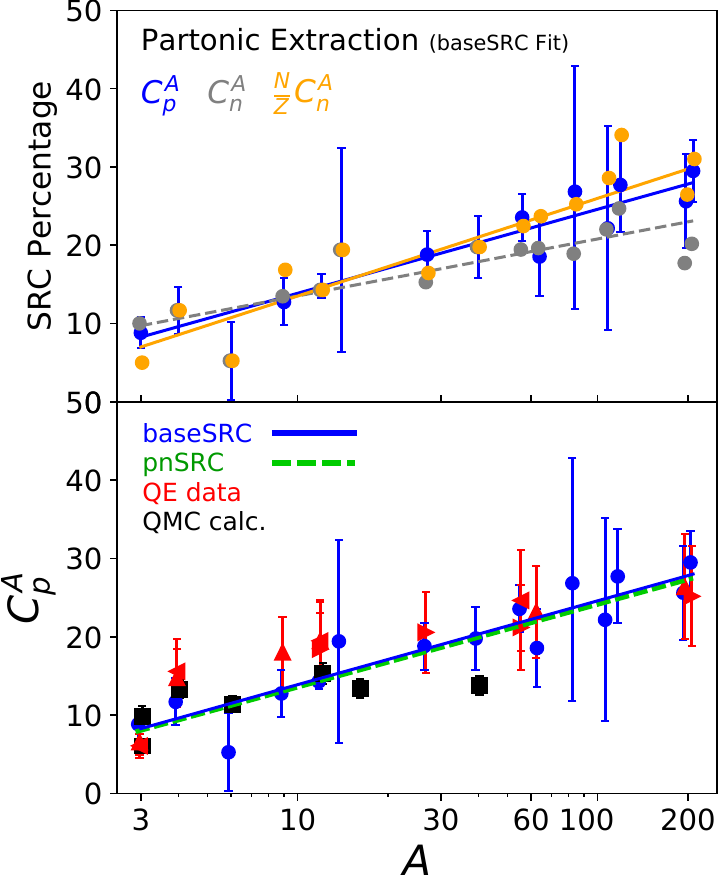}
\caption{(a)~Comparison of nuclear structure parameters $C_p^A$, $C_n^A$, and $(N/Z)C_n^A$ values for the {\tt baseSRC} fit. 
The solid lines represent logarithmic fits to the corresponding quantities.
We show uncertainties only for the $C_p^A$, 
but errors for other quantities are of similar size.
(b)~Comparison of $C_p^A$ values for  the {\tt baseSRC} fit 
and the  SRC abundances extracted from quasi-elastic~(QE)~\cite{CLAS:2005ola,Fomin:2011ng,Schmookler:2019nvf} data
and  Quantum Monte Carlo (QMC)~\cite{Cruz-Torres:2019fum} nuclear calculations. 
The logarithmic fits for  {\tt baseSRC} and {\tt pnSRC} are also shown.
\\
}
\label{fig:one}
\end{figure}

We keep a model-independent
approach as to the number of SRC pairs and their isospin structure. In fact, these nuclear structure parameters will be
independently determined in our nPDF analysis for the first time ({\it cf.},~Fig.\,\ref{fig:one}), and tested for consistency with independent results
from specific nuclear structure studies.

We further note that Eq.\,(\ref{eq:4fSRC})  represents a natural evolution of previous studies that: (i)~observed a linear
correlation between measured SRC abundances and the modified structure of bound nucleons in the valence region (i.e.,
at high fractional-momentum ${x \sim 0.3 {-} 0.7}$), known as the EMC effect~\cite{Weinstein:2010rt,Hen:2012fm,Hen:2013oha}, and (ii)~showed that this correlation can
result from a universal modification of the valence region structure function of nucleons in proton-neutron SRC pairs~\cite{Schmookler:2019nvf}.
These studies were limited to the nucleon valence region (${x \sim 0.3 {-} 0.7}$) and using lepton-nucleus Deep
Inelastic Scattering (DIS) data with specific nuclear structure input for SRC abundances~\cite{Schmookler:2019nvf,Segarra:2019gbp}. 
As detailed below, here we
study the fundamental nPDFs, extend the analysis to the full $x$~range of $10^{-3}$ to 0.8, use a comprehensive 
 data set on the parton structure of nuclei, and do not impose external inputs for nuclear structure parameters  $C^A_N$. 
The latter is especially important as the independent extraction of  $C^A_N$ allows us to compare with known
measured and calculated nuclear structure values to support, or challenge, the physical validity and interpretation of
Eq.\,(\ref{eq:4fSRC}).

As the DIS coherence length scales as $1/x$,
it is more natural to associate  {higher-$x$} phenomena with short-range nuclear physics 
than very {low-$x$}
(${x \sim 10^{-2} {-} 10^{-3}}$) phenomena.
Low-$x$ reactions are sensitive to the number of nucleons the virtual photon propagates through, 
given by the nuclear thickness function~\cite{Alvioli:2009iw}:
\begin{eqnarray}
  T^{(2)}(b) &=&
  \int dz_1 \int dz_2 \,\,\rho^{(2)}(b_1{=}b,z_1;b_2{=}b,z_2) \ ,
\end{eqnarray}
where $\rho^{(2)}$ is the two-nucleon density defining the probability for finding two nucleons with transverse
and longitudinal positions $b$ and $z$. 
For \mbox{mean-field} models, we can assume $\rho^{(2)}$ approximately factorizes into 
${\sim} \rho(b_1,z_1)\rho(b_2,z_2)$.
Studies show the SRC pairs significantly impact the two-nucleon density leading to a
typical correction of~\cite{Miller:1975hu,Cruz-Torres:2017sjy}
\begin{align}
  \rho^{(2)}(b_1,z_1;&b_2,z_2) \approx \nonumber\\
  \rho(&b_1,z_1) \rho(b_2,z_2) \left\{1+C\left( |z_1-z_2| \right) \right\} \ .
\end{align}
Here, the correlation function 
$C(|z_1-z_2|)$ is sensitive to the number of nucleons in SRC pairs 
({\it i.e.},  $C^A_N$).
Therefore,  $\rho^{(2)}$  modifies the  nuclear thickness function, which will impact 
the low-$x$ region. 
This does not mean to imply that the measured behavior at low-$x$
stems only from the modification of the structure of nucleons in SRC pairs,
but it can potentially depend on the SRC-induced correlation. 
This requires further investigation with additional low-$x$ data.

\sectionX{Analysis and Results}
The analysis utilized the full set of available world data on nuclear lepton DIS, Drell-Yan processes, and $W$ and $Z$ boson
production (see the Supplementary Material for data-set details).
The corresponding theory predictions were obtained in collinear factorization with the parton-level cross-sections calculated at the next-to-leading order (NLO) of QCD.
The DIS data primarily constrain the $u$ and $d$ quark and anti-quark distributions, whereas the $W$ and $Z$ boson production data from LHC proton-lead collisions also constrain
strange quark and gluon distributions down to lower momentum fractions of $x\,{\sim}\,10^{-3}$.

Energy scale dependence is accounted for using the DGLAP evolution equation, which also helps constrain
the gluon distribution through the $Q$-dependence of DIS data. Parton number and momentum sum rules are ensured to be
separately satisfied for $f_i^{p(n)}(x,Q)$ and 
$f_i^{\mathrm{SRC}\, p(n)}(x,Q)$, and are therefore also satisfied for their linear combination in
Eq.~\eqref{eq:4fSRC}
independently of the values of  $C_p^A$ and  $C_n^A$.%
\footnote{Note that the sum rules for the free-nucleon distributions are satisfied by construction. This is ensured by relying on the free-nucleon PDFs determined in a dedicated fit~\cite{Stump:2003yu}.}

The free-nucleon PDFs  $f_i^{p(n)}(x,Q)$ are fixed to the distributions determined in Ref.\,\cite{Kovarik:2015cma}
via global analysis of nucleon structure observables. The SRC nucleon PDFs 
$f_i^{\mathrm{SRC}\, p(n)}(x,Q)$ use the same functional form as 
$f_i^{p(n)}(x,Q)$ with 21 shape parameters that are fit herein. We perform two independent
analyses where i)~$C^A_p$ and  $C^A_n$ are allowed to vary freely, and ii)~where we assume proton-neutron SRC dominance,
i.e.~$C^A_p=\frac N Z \times C^A_n \equiv C^A$. We refer to these fits as the {\tt baseSRC} and {\tt pnSRC} fits, respectively. 
For comparison, we also repeated the traditional (mean-field-like) analysis of Ref.\,\cite{Segarra:2020gtj} using the same dataset used herein, which we refer to as the \traditional\ fit.

 The resulting fit quality in terms of  $\chi ^2$ for each SRC fit and for the \traditional\ fit are
listed in Table~\ref{tab:fits} for each data type separately, and for all data combined.
As can be seen, the SRC fit 
using  Eq.\,(\ref{eq:4fSRC}) results in overall  
$\chi_{{\rm tot}}^{2}/N_{\rm DOF}$ values  
appreciably better than for the \traditional\ fit. 
This takes into account the additional SRC ($C_{p,n}^A$) 
parameters. At the level of the total $\chi^2$ we obtain 
a reduction of 108 and 58 points for the \baseSRC{} and 
the \pnSRC{} fits respectively.
We find that the nPDFs for the \traditional{} and SRC  fits are in general agreement within uncertainties.
All data are well reproduced for the full range of the data, 
corresponding to an $x$ range of about $10^{-3}$ to~0.75.
\begin{table}[tb]
  \centering
  \begin{tabular}{|c||c|c|c|c||c||c|}
    \hline   %
    $\chidat$ & DIS  & DY  & $W/Z$  & JLab  & 
    $\chi_{{\rm tot}}^{2}$  
    &
    {\scriptsize \small $\frac{\chi_{{\rm tot}}^{2}}{N_{\rm DOF} }$  }%
    \tabularnewline[4pt]
    \hline 
    \hline    %
    \traditional & 0.85 & 0.97 & 0.88 & 0.72 &  1408 & 0.85
    \tabularnewline
    \hline    %
    \baseSRC   & 0.84 & 0.75 & 1.11 & 0.41 & 1300  &  0.80
    \tabularnewline
    \hline    %
    \pnSRC   & 0.85 & 0.84 & 1.14 & 0.49 & 1350 & 0.82
    \tabularnewline
    \hline 
    \hline    %
    $N_{\rm data}$ & 1136 & 92 & 120 & 336 & 1684 & 
    \tabularnewline
    \hline    %
  \end{tabular}
  \caption{
  The partial $\chidat$ values for the data subsets; 
  the number of data points ($N_{\rm data}$) for each process
  are listed in the bottom row. 
  The \traditional\ fit has 19~shape and 3~$W/Z$ normalization parameters.
  The \baseSRC\ and \pnSRC\ fits have 21~shape, 3~$W/Z$ normalization, and 30~and 19~SRC parameters ($C^{p,n}_A$), respectively.
  In total, there are 1684 data points after cuts. 
  Note the last column ($\chi_{{\rm tot}}^2/N_{\rm DOF}$) fully takes into account the number of fit parameters.
  }
  \label{tab:fits}
\end{table}

Figure\,\ref{fig:one} shows the extracted  $C_p^A$ and  $C_n^A$ coefficients as determined by the global {\tt baseSRC}.
The coefficients show logarithmic growth with the nuclear mass number~$A$, starting from ${\sim}5\%$ for helium-3 and reaching
${\sim20}\%$  for lead. 
As shown in \mbox{Fig.\,\ref{fig:one}-b)},  
the {\tt baseSRC} and {\tt pnSRC} fits give similar results with the {\tt baseSRC} fit preferring a similar number of SRC protons and neutrons, even for heavy neutron-rich nuclei. This dynamics is consistent with the observation of \mbox{$pn$-dominance}, previously determined from nuclear structure studies~\hbox{\cite{Subedi:2008zz,Piasetzky:2006ai}.} 
Therefore, the {\tt baseSRC} and {\tt pnSRC} consistency is a first indication of consistency between quark-gluon level analysis and nuclear structure studies.

Focusing on the lower panel, Figure\,\ref{fig:one} also shows the extracted  $C_p^A$ 
coefficients are consistent with previous, independent, extractions from
both nuclear structure calculations~\cite{Cruz-Torres:2019fum} and quasi-elastic electron scattering measurements~\cite{CLAS:2005ola,Fomin:2011ng,Schmookler:2019nvf}. 
As the nuclear structure
calculations assume point-like nucleons, and the measurements are done at energies that only probe nucleons (and not their substructure), they are both insensitive to the nucleon nPDFs. 

Therefore, we report here the first direct extraction of nuclear structure information 
from parton-level observables in an nPDF data analysis that is fully consistent with independent nuclear structure extractions.

\begin{figure}[tb]
\centering
\includegraphics[width=0.45\textwidth]{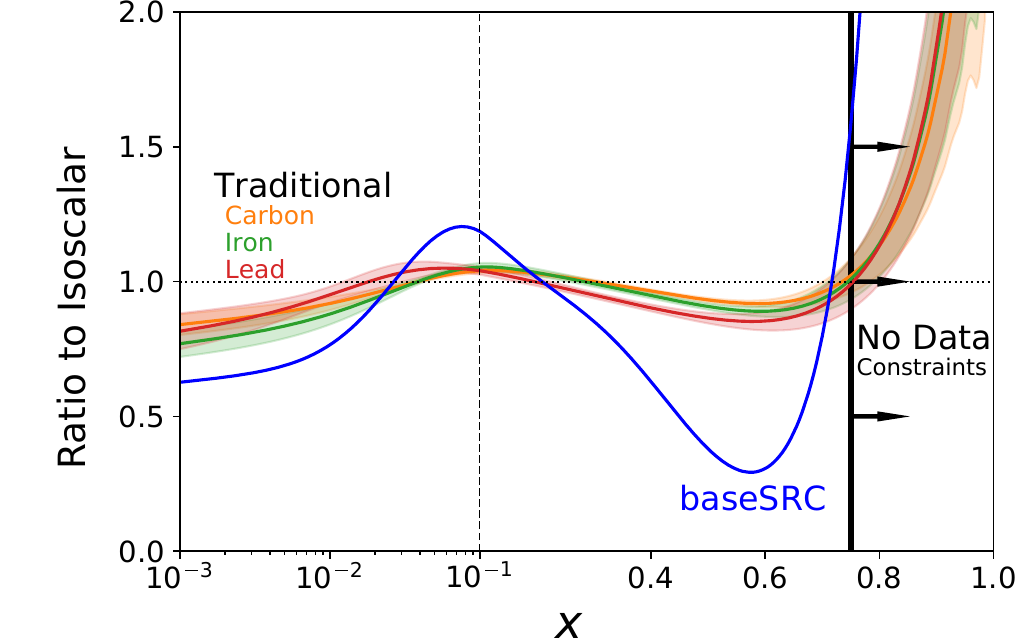}
\caption{
The ratio of the rescaled structure function $F_2^A/A$ to the isoscalar combination \mbox{$(F_2^p{+}F_2^n)/2$}, computed for the \traditional\ PDFs for carbon, iron and lead. Separately, we show the isoscalar $F_2$ structure function computed with the SRC component, $f_i^{\text{SRC}}$, of the \baseSRC\ PDFs divided by the aforementioned isoscalar combination. 
Both $F_2^A$ and $F_2$ are calculated using the LO formula~\cite{ParticleDataGroup:2020ssz} at $Q{=}10$~GeV. 
The~\baseSRC\ curve illustrates the shape of the relative nuclear modification, which is universal and independent of $A$.
This nuclear modification is weighted by the SRC coefficients 
(typically $\sim$10\% to 30\%) and added to the proton PDF to yield the full nPDF.
}
\label{fig:fModpn}
\end{figure}

Finally, looking at the nucleon structure itself, Fig.\,\ref{fig:fModpn} shows the 
ratios of rescaled structure functions $F_2^A /A$, compared to the \mbox{$F_2$-isoscalar}  ({\it i.e.}, \mbox{$(F_2^p{+}F_2^n)/2$}) for the \traditional\ fit. 
We also compute the $F_2$ structure function ratio 
for nucleons in SRC pairs relative  to \mbox{$F_2$-isoscalar};
results shown are obtained from the {\tt baseSRC} fit (blue curve). 
Direct measurements of the ratio $F_2^A/A$  to $F_2$-isoscalar exist for the low $A$ deuteron and triton measurements~\cite{Griffioen:2015hxa,Li:2022fhh} 
that are not included in the present analysis; 
extending the current analysis to lighter nuclei, including the mirror nuclei ${}^3\mathrm{H}$ and ${}^3\mathrm{He}$, 
are planned for future study.

As can be seen in  Fig.\,\ref{fig:fModpn},
the SRC and \traditional\ curves both result in  
qualitatively similar deviations from the free-nucleon structure, with the 
SRC  showing significantly  larger modification than even lead (Pb)
in the \traditional\  case.
This is expected as the SRC curve corresponds only to the one component of full distribution (c.f.~Eq.~\eqref{eq:4fSRC}), which is responsible for the whole nuclear modification. 
The amount by which the SRC modification is larger than the \traditional\  case, however, varies significantly with $x$.
At very {low-$x$}, below ${ 2 - 3 \times 10^{-2}}$, we observe only a slightly increased suppression for the SRC case.
At intermediate-$x$ of ${2 - 3 \times 10^{-2}}$ to 
${2 \times 10^{-1}}$ we observe a more pronounced difference between the SRC and \traditional\  enhancements.
Furthermore, at high-$x$ value, above ${\sim}0.6$, the difference is most pronounced.
Part of this enhanced high-$x$ effect can be understood to result from the effects of Fermi-motion that are known to grow with $x$ and are included
in our approach inside the modification function and therefore leads to the appearance of enhanced modification effects at \mbox{high-$x$~\cite{Segarra:2020plg}.}

As noted earlier,  when we combine the elements of Fig.\,\ref{fig:fModpn} to construct the nPDFs, we find that the valence nPDFs for the \traditional\ and SRC  fits are identical, within uncertainties.
This highlights the fact that the nPDFs are truly constrained by the data, despite the differing parameterizations.

\sectionX{Conclusions}
We have performed the first-ever global QCD analysis of nuclear PDFs using a framework based on concepts from SRC nuclear models. It leads to similar, or better, data description as compared to the traditional parameterization, and enables a meaningful physical interpretation of the fit. The incorporated data include the high-energy DIS, DY, and electroweak boson production commonly used in nPDF fits. The analysis determines both the standard ``average" nuclear PDFs (that can be compared with traditional nPDF fits), as well as a universal distribution of partons in SRC nucleon pairs and the fractions of such SRC pairs.

This analysis represents a direct extraction of nuclear structure information from experimental observables directly probing quark-gluon nuclei dynamics. The fact that the obtained fractions of SRC pairs agree with their previous extractions from the low-energy quasielastic data establishes a direct link between high-energy partonic properties and lower-energy nuclear physics. It thus presents a significant advance in our quest to understand atomic nuclei in terms of QCD. Furthermore, the extracted distributions of partons in SRC pairs can be directly tested using measurements of tagged processes at the Jefferson Lab accelerator and the future Electron-Ion Collider.

This new nPDF set can also potentially impact the analysis of heavy-ion measurements that require a combination of nuclear PDFs, together with initial state nuclear matter effects~\cite{Ferreiro:2008wc,Ferreiro:2009ur,Ferreiro:2012sy}. 
 Whereas traditional approaches thus far assign the same nPDF to all nucleons in the calculated initial-state distributions, the SRC approach allows additional flexibility. With the SRC PDFs, we can i) follow the traditional approach and simply use averaged distributions, or ii) we can construct a more complex initial-state nucleon distribution using a combination of the free-nucleon PDF and SRC-modified PDF to each nucleon depending on its correlation state.
Furthermore, it is noteworthy that the SRC parameterization (in which the dependence of $A$ and $x$ is factorized) produces an excellent description of the data; 
the conceptual simplicity of this parameterization is striking. 

\appendix
\onecolumngrid
\vspace{1.5cm}
\twocolumngrid
\section*{Supplementary Material}

\subsectionX{The Data} %
For this analysis, we used data from nuclear DIS (including JLab), Drell-Yan (DY), 
and high-mass $W/Z$ production.
These are the same data as used in the nCTEQ15HIX analysis of Ref.~\cite{Segarra:2020gtj}
with the addition of the LHC $W/Z$ production data from proton-lead collisions used in
the nCTEQ15WZ~\cite{Kusina:2020lyz} analysis.
As with nCTEQ15HIX, we choose kinematic selection cuts of $Q{>}1.3$~GeV and $W{>}1.7$~GeV
for DIS data. These cuts are relaxed as compared to the usual requirements
($Q{>}2$~GeV and $W{>}3.5$~GeV) enforced in previous nCTEQ nPDF analyses, which allows us to include more
of the high-$x$ data in particular data from JLab experiments. 

\subsectionX{The Fits} %
For our fit to the nuclear data, we fix the proton PDFs $f_{i/p}(x,Q)$ 
to the nCTEQ15 proton distributions~\cite{Kovarik:2015cma}.
We have also explored other proton sets from different groups,
e.g.~\cite{Hou:2019efy,H1:2015ubc,Accardi:2016qay,NNPDF:2017mvq}
and obtained comparable results.

We parameterize the SRC PDFs using a similar form as in both nCTEQ15 and nCTEQ15HIX:
\begin{eqnarray}
    \label{eq:parm}
    x f_i^{SRC}(x,Q_0) &=&
    p_0 x^{p_1} (1-x)^{p_2} e^{p_3 x} (1+ e^{p_4} x)^{p_5}\  ,\\
    \frac{\bar{d}^{SRC}}{\bar{u}^{SRC}}(x,Q_0) &=& p_0x^{p_1}(1-x)^{p_2} + (1+p_3x)(1-x)^{p_4}\  .
    \nonumber
\end{eqnarray}
Specifically, we parameterize the following flavor combinations:
$\{u_v,d_v,(\bar{u}{+}\bar{d}),(\bar{d}/\bar{u}),(s{+}\bar{s}),g\}$ 
and 21 shape parameters:
$\{ 
p_{1...5}^{u_v}, 
p_{1,2,3}^{d_v}, 
p_{1,2,3}^{\bar{u}+\bar{d}}, 
p_{0...5}^{g}, 
p_{0,1,2}^{s{+}\bar{s}}, 
p_{5}^{\bar{d}/\bar{u}}
\}.$
The free-nucleon component, $f^{p(n)}(x,Q)$, is not fitted and is instead adapted from a dedicated global analysis~\cite{Stump:2003yu}.

Additionally, for each non-isoscalar nucleus we fit the $C_p^A$ and $C_n^A$
coefficients from Eq.\,\eqref{eq:4fSRC}, which count the fractions of protons
and neutrons in the SRC pairs. For isoscalar nuclei we set $C_p^A=C_n^A$.
The data used in our analysis covers 18 nuclear targets:
\{${}^{2}$H,
${}^{3}$He, 
${}^{4}$He, 
${}^{6}$Li, 
${}^{9}$Be$_{\rm iso}$, 
${}^{12}$C, 
${}^{14}$N, 
${}^{27}$Al, 
${}^{40}$Ca, 
${}^{56}$Fe, 
${}^{64}$Cu, 
${}^{84}$Kr, 
${}^{108}$Ag, 
${}^{119}$Sn, 
${}^{131}$Xe, 
${}^{184}$W, 
${}^{197}$Au, 
${}^{208}$Pb\}
with 6 of them:
$\{ \rm {}^{2}H, {}^{4}He, {}^{6}Li, {}^{9}Be_{iso}, {}^{12}C, {}^{40}Ca\}$,  
being isoscalar (or isoscalar corrected).
This yields a total of 30 parameters controlling the SRC fractions for the {\tt baseSRC} fit,
and 19 for the {\tt pnSRC} fit.
For very weakly constrained nuclei (${}^{131}$Xe, ${}^{184}$W) we include additional bounds on $C_{p(n)}^A$ coefficients. 
In the \pnSRC\ fit we also keep the $C_p^A$ and $C_n^A$ free for ${}^3$He. 

The {\tt traditional} fit uses the same parametrization and open parameters as the nCTEQ15HIX analysis~\cite{Segarra:2019gbp}. Each flavour combination of the total nuclear PDF is  at the input scale parameterized using the same functional form as in Eq.\,\eqref{eq:parm}, with the $p$ parameters replaced by
\begin{align}
    p_k \longrightarrow p_k(A) \equiv p_{0,k}+ a_{k}(1-A^{-b_{k}}) \ ,
\end{align}
to encode the nuclear $A$ dependence.
The parameters $p_{0,k}$, i.e.~the free-nucleon component, are kept fixed just like in the SRC fits.

To summarize the complete set of  fitting parameters, 
the \traditional\ fit has 19~shape and 3~$W/Z$ normalization parameters. 
The 
\baseSRC\ and \pnSRC\ fits  use an invariant ``free-nucleon" PDF $f_i^{p,n}(x,Q)$,
21~shape parameters for the SRC PDF, with 3~$W/Z$ normalization, and 30~and 19~SRC parameters ($C^{p,n}_A$), respectively. 
In total, there are 1684 data points after cuts, and we 
fully take into account the complete set of fitting parameters when computing 
$\chi_{{\rm tot}}^2/N_{\rm DOF}$ in Table~\ref{tab:fits}.
For comparison, EPPS21~\cite{Eskola:2021nhw} use a total of 24 parameters,
and nNNPDF3.0~\cite{AbdulKhalek:2022fyi} uses a neural network model with on order 200 parameters.

For our analysis, we use the  \ncteqpp{} framework, which has a modular structure
and links to a variety of external tools, including
a modified version of HOPPET~\cite{Salam:2008qg}
(extended to accommodate grids of multiple nuclei),
APPLgrid~\cite{Carli:2010rw}, and MCFM~\cite{Campbell:2015qma}.
Additionally, we have used FEWZ~\cite{Li:2012wna} for benchmarking
our $W/Z$ calculations,
and xFitter~\cite{Alekhin:2014irh} for various cross-checks.

\null
\vspace{-1.0cm}
\section*{Acknowledgments}
We are grateful to 
Jeff Owens  
and 
Efrain Segarra
for valuable discussion.
The work at University of M{\"u}nster was funded by the DFG through the Research Training Group 2149 ``Strong and Weak Interactions--from Hadrons to Dark Matter'' and the SFB 1225 ``Isoquant,'' {project\nobreakdash-id} 273811115. 
The research of P.~D.\ was funded as a part of the Center of Excellence in Quark Matter of the Academy of Finland (project 346326).
M.K.~thanks the School of Physics at the University of New South Wales in Sydney, Australia for its hospitality and financial support through the Gordon Godfrey visitors program.
A.K.~and N.D.~are grateful for the support of Narodowe Centrum Nauki under grant no.\ 2019/34/E/ST2/00186.
The work of  C.K.~was supported by the  U.S.\ DoE contract DE-AC05-06OR23177, under which Jefferson Science Associates LLC manages and operates Jefferson Lab. 
J.G.M. has been supported by Fermi Research Alliance, LLC under Contract No. DE-AC02-07CH11359 with the U.S. DoE.
F.O.~was supported by the U.S.~DoE under Grant No.~DE-SC0010129, 
and  the DOE Nuclear Physics, within the Saturated Glue (SURGE) Topical Theory Collaboration.
R.R.~is supported by the Polska Akademia Nauk (grant agreement PAN.BFD.S.BDN.\ 613.022.2021-PASIFIC 1, POPSICLE) and by Narodowe Centrum Nauki under Grant No. 2023/49/B/ST2/04330. This work has received funding from the European Union's Horizon 2020 research and innovation program under the Sk{\l}odowska-Curie grant agreement No.~847639 and from the Polish Ministry of Education and Science.
The work of I.S.\ was supported by the French CNRS via the IN2P3 project GLUE@NLO.
The work of E.P.\ was supported by the OPRA foundation, Pazy foundation, and the Israeli Science Foundation (Grant 917/20).

\bibliographystyle{apsrev4-1}
\bibliography{main}
\end{document}